\documentclass[twocolumn]{aastex62}

\usepackage{amssymb}
\usepackage{amsmath}
\usepackage{amssymb}
\usepackage{empheq}
\usepackage{mathrsfs}
\usepackage{wasysym} 

\usepackage{etoolbox}
\usepackage{lmodern}

\newcommand{\be}{\begin{equation}}
\newcommand{\ee}{\end{equation}}
\newcommand{\G}{\mathcal{G}} 
\newcommand{\M}{M_{\star}} 
\newcommand{\B}{B_{\star}}
\newcommand{\R}{R_{\star}}
\newcommand{\Mdot}{\dot{M}}

\def\lta{\,\raise 0.3 ex\hbox{$ < $}\kern -0.75 em
 \lower 0.7 ex\hbox{$\sim$}\,}
\def\gta{\,\raise 0.3 ex\hbox{$ > $}\kern -0.75 em
 \lower 0.7 ex\hbox{$\sim$}\,}

\def\lta{\,\raise 0.3 ex\hbox{$ < $}\kern -0.75 em
 \lower 0.7 ex\hbox{$\sim$}\,}
\def\gta{\,\raise 0.3 ex\hbox{$ > $}\kern -0.75 em
 \lower 0.7 ex\hbox{$\sim$}\,} 
\newcommand{\rhobar}{\langle\rho\rangle} 

\begin{document} 

\title{\bf The Origin of Universality in the Inner Edges of Planetary Systems} 


\author[0000-0002-7094-7908]{Konstantin Batygin}
\affiliation{Division of Geological and Planetary Sciences, California Institute of Technology, Pasadena, CA 91125}

\author[0000-0002-8167-1767]{Fred~C.~Adams}
\affiliation{Physics Department, University of Michigan, Ann Arbor, MI 48109}
\affiliation{Astronomy Department, University of Michigan, Ann Arbor, MI 48109}

\author[0000-0002-7733-4522]{Juliette Becker}
\affiliation{Division of Geological and Planetary Sciences, California Institute of Technology, Pasadena, CA 91125}




\begin{abstract}
The characteristic orbital period of the inner-most objects within the galactic census of planetary and satellite systems appears to be nearly universal, with $P$ on the order of a few days. This paper presents a theoretical framework that provides a simple explanation for this phenomenon. By considering the interplay between disk accretion, magnetic field generation by convective dynamos, and Kelvin-Helmholtz contraction, we derive an expression for the magnetospheric truncation radius in astrophysical disks, and find that the corresponding orbital frequency is independent of the mass of the host body. Our analysis demonstrates that this characteristic frequency corresponds to a period of $P\sim3$  days, although intrinsic variations in system parameters are expected to introduce a factor of $\sim2-3$ spread in this result. Standard theory of orbital migration further suggests that planets should stabilize at an orbital period that exceeds disk truncation by a small margin. Cumulatively, our findings predict that the periods of close-in bodies should span $P\sim2-12$ days -- a range that is consistent with observations.

\end{abstract}


\section{Introduction}
\label{sec:intro} 

Planetary systems and satellites of gas giants are widely believed to form within moderately massive circumstellar and circumplanetary disks, which are composed primarily of Hydrogen and Helium. Although the exact mode of their growth depends on the local nebular conditions \citep{Safronov1972,Lissauer1993,Lambrechts2012,BM2023}, the accretion process -- if fast enough -- unavoidably triggers some degree of orbital migration \citep{Ward1997}. After becoming massive enough to raise substantial wakes within their parent nebulae, planets and satellites begin to exchange energy and angular momentum with the surrounding gas (e.g., see  \citealt{Paardekooper2022} for a recent review). This exchange is irreversible, and typically results in orbital decay, which drives the forming objects towards the inner edge of the disk. Near this edge, which is characterized by a sharp decline in the nebular surface density, the migratory torques reverse sign so that inward migration comes to a halt \citep{Masset2006}.  

\begin{figure}[t]
    \centering
    \includegraphics[width=\linewidth]{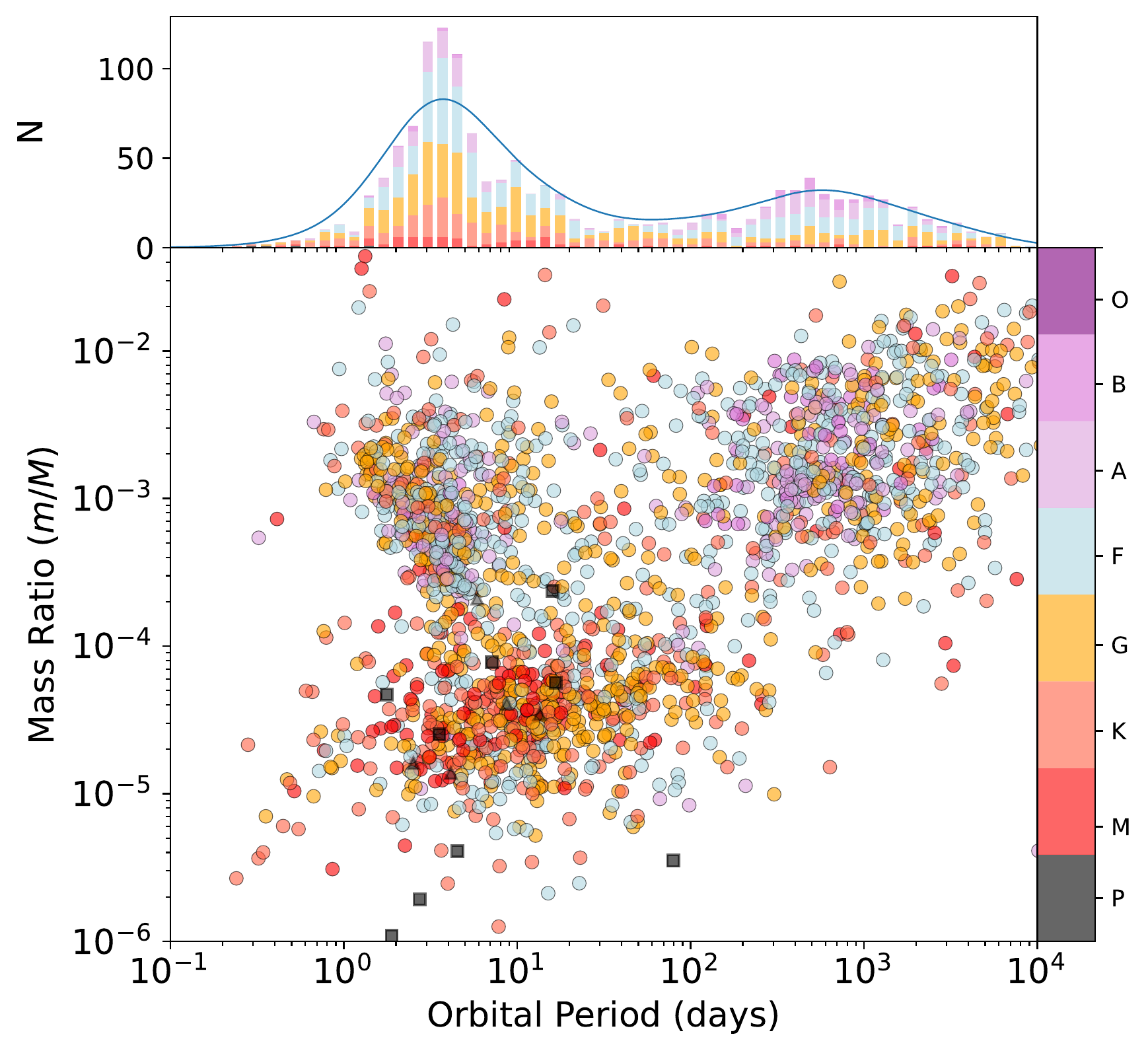}
    \caption{Census of well-characterized extrasolar planets (circles) and moons of gas (squares) and ice (triangles) giants of the solar system. The mass ratio of the planet or satellite to the central body ($m/M$) is shown as a function of the orbital period. The points are color-coded by the spectral type of the host (with P denoting a planetary central body). The top panel shows a histogram of the orbital period of the innermost planet or satellite in each system, similarly color-coded by the spectral type of the host star. A peak in the $P\sim2-12$ day range is evident within the distribution. A complementary view of the orbital distribution is provided in Figure \ref{fig:panels}.}
    \label{fig:data}
\end{figure}

The inner cavities of astrophysical disks are carved out by the magnetospheres of their central objects, such that the truncation radius is determined by a balance between Lorentz torques and viscous torques, where the latter facilitate gas accretion  \citep{Koenigl1991,Bouvier2007,Batygin2018}. As a result, the present-day orbits of short-period planets and satellites must, to some extent, reflect this primordial equilibrium (e.g., \citealt{lee2017}). Given that the masses, magnetic moments, and mass-accretion rates of central objects can vary dramatically from system to system, it comes as no surprise that the semi-major axes of the inner-most orbits within the galactic census of planets and moons shows enormous variability. On the other hand, the corresponding orbital periods display a striking pattern of uniformity, as shown in Figure \ref{fig:data} (see also Figure \ref{fig:panels}). Specific examples of this pattern include the orbital periods of Io, Trappist-1b, and Kepler-11b (as well as many others), which are all measured to be several days, despite the fact that the masses of their parent bodies span three orders of magnitude. 


Beyond the cursory view of the data provided by Figure \ref{fig:data}, the inner edges of planetary systems have been quantified in a statistically rigorous manner by \citet{Petigura2018, Mulders2018}. For the Super-Earth population, \citet{Petigura2018} find a knee in the planet occurrence distribution at a period of $P \sim 6$ days, while \citet{Mulders2018} derive a distribution for the inner-most orbital periods within the Kepler sample that exhibits a flattened peak between 2 and 8 days, and drops off rapidly outside of these bounds. In the same vein, it is worth noting that the knee in the occurrence distribution of sub-Neptunes\footnote{Following the work of \citet{Fulton2017}, \citet{Petigura2018} define sub-Neptunes as objects with a radius greater than $1.7\,R_{\oplus}$.} ensues at a larger period (i.e., $P \sim 12$ days; \citealt{Petigura2018}) than that of the Super-Earths. Though interesting, this difference likely stems from instability of H/He envelopes to evaporation at shorter orbital periods \citep{Rogers2021}, meaning that the characteristic value corresponding to the Super-Earth population is likely to be more representative of the primordial inner edges of planetary systems. At the same time, we acknowledge that the dataset for planets with radii smaller than Earth remains highly incomplete. Yet, we expect that the reversal of type-I migration at the disk cavity would operate similarly for super-Earths and sub-Terrestrial planets \citep{Masset2006}, and for this reason, we assume that the orbital distribution of undetected planets is similar to that of the observed ones.

The case of short-period giant planets is somewhat more complex. While a power-law occurrence distribution does not provide an adequate match to the data, it is important to understand that giant planets are expected to appreciably modify the structure of the disk through gap-opening. Given the complexity of type-II migration in 3D, it is not clear that Jovian-class objects must systematically migrate all the way down to the magnetospheric cavities, as in the case of lower-mass bodies that migrate in the type-I regime \citep{Lega2022}. Nevertheless, previous work has shown that the inner boundary of the hot Jupiter population can be fully explained by magnetic disk truncation, and the characteristic $r\propto m^{2/7}$ relationship of the dividing line emerges self-consistently from consideration of gas accretion and the appearance of a magnetospheric cavity within the nebula \citep{Bailey2018}. An additional complication, however, lies in that an unknown fraction of the hot Jupiter population migrated inward through the high-eccentricity channel, and the $r \sim m^{1/3}$ relationship expected within the context of this picture is also consistent with the data \citep{Matsakos2016,Owen2018}.

Finally, within the context of planetary satellites, our analysis is limited to large natural satellites of giant planets that formed within primordial circumplanetary disks. Moons that originated from giant impacts, capture, or from rings are beyond the scope of this work\footnote{Our work therefore excludes the large satellites of Uranus and Neptune — which likely formed through a giant impact \citep{Ida2020} and capture \citep{Agnor2006}, respectively.}. Correspondingly, the minuscule Jovian moons such as Metis, Adrastea, Amalthea, and Thebe represent a separate category of objects when compared to the massive Galilean moons. Similarly, there is no reason to suspect that the small moons in the Saturnian system — which likely formed from the rings and experienced tidal migration outwards \citep{Crida2013} — should adhere to the magnetospheric edge of Saturn’s circumplanetary disk, since they emerged after the dissipation of the gas. As for Titan, it is worth noting that while its current orbital period is approximately two weeks, recent work by \citet{Lainey2020} convincingly demonstrates that Titan has undergone rapid orbital expansion, with evidence indicating that its primordial period was on the order of a few days.

\begin{figure}[t]
    \centering
    \includegraphics[width=\linewidth]{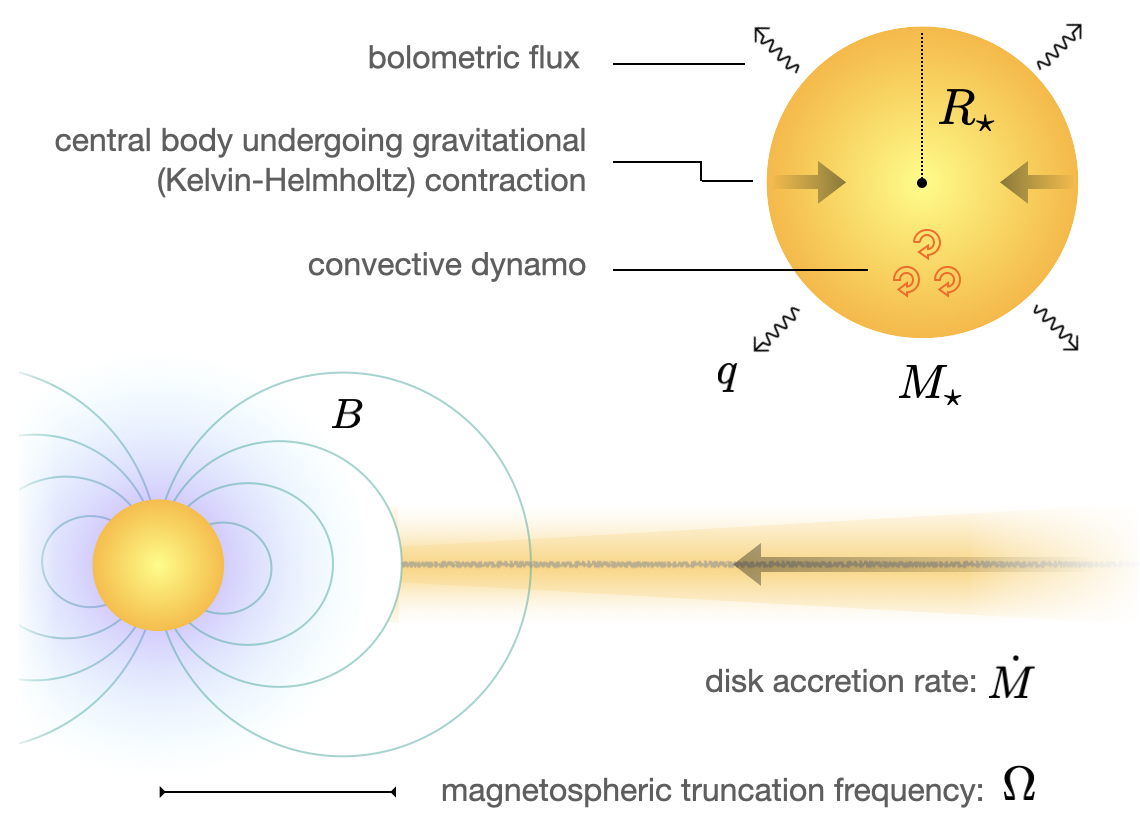}
    \caption{A schematic diagram of the physical processes considered in this work. A central body of mass $\M$ is encircled by a disk, which accretes at a rate $\Mdot$. Convective energy transport within the body drives gravitational contraction of its radius, $\R$, as well as dynamo action, which generates a field, $\B$, that is regulated by the bolometric flux, $q$. Magnetospheric interactions between the central body and the disk generate a cavity with an edge that corresponds to an orbital frequency, $\Omega$.}
    \label{fig:diagram}
\end{figure}

The principal goal of this letter is to explain this near-universality of the characteristic orbital period associated with magnetospheric disk truncation. Toward this end, we construct an analytic framework that draws upon conventional results from astrophysical dynamo theory, stellar structure, and disk accretion. The relevant physical processes are summarized in Figure \ref{fig:diagram}. Taken together, these considerations reveal how planetary architectures are assembled within their natal nebulae, and shed light on the fundamental role played by magnetospheric cavities in shaping the outlines of the inner boundaries of the resulting planetary systems. 

\section{Analytic Theory}
\label{sec:argument} 

As a starting point for our analysis, let us recall the standard expression for magnetospheric truncation of an astrophysical disk. Though formally incorrect, this condition is routinely derived by equating the magnetic pressure, $B^2/(2\,\mu_0)$, to the accretionary ram pressure, $\rho\,v_r^2$ (e.g., \citealt{MohantyShu2008}), with the result 
\begin{align}
\frac{B^2}{2\,\mu_0} \sim \frac{\dot{M}}{4\,\pi\,r^2}\sqrt{\frac{2\,\G\,\M}{r}},
\end{align}
where we have written the mass-accretion rate of the disk as $\Mdot = 4\pi\,r^2\,\rho\,v_r$, appropriate for spherical infall. Assuming a disk-aligned dipole field for the star $B=\B\,(\R/r)^3$, this expression can be solved for the orbital radius $r$ (e.g., \citealt{GhoshLamb1978,Blandford1982}). In a slight variation on this theme, here we write the truncation condition not in terms of the orbital radius, but rather in terms of the orbital frequency, 
\begin{align}
\Omega = \xi \, \bigg( \frac{\mu_0^3\,\G^5\,\M^5\,\Mdot^3}{2\,\sqrt{2}\,\pi^3\,\B^6\,\R^{18}} \bigg)^{1/7},
\label{omegatr}
\end{align}
where $\xi$ is a proportionality constant of order unity\footnote{The exact value of this constant depends on the field geometry. As an example, detailed dipole calculations of \citet{NajitaShu1994} yield $\xi\approx1.13$.}. In the following discussion, we examine how each of the physical quantities inside the parenthesis depend on the mass of the central object (and other properties of the system). 

\paragraph{Mass-Accretion Rate} At the detailed level, nebular accretion is a complex and episodic process that is facilitated by turbulent viscosity and magnetized disk winds (e.g., see  \citealt{ArmitageBook2020,ArmitageAccretionNotes} and references therein). Nevertheless, the overall magnitude of $\Mdot$ is empirically known to increase with the mass of the central object and to diminish with time \citep{Hillenbrand1992,Hartmann1998,Herczeg2008}. Moreover, gravitational stability regulates the ratio $\beta$ of the initial disk mass to the central object mass to be of order 1/10 \citep{Shu1990}, i.e., comparable to its geometric aspect ratio so that $\beta\sim0.1\sim h/r$. The simplest parameterization that we can adopt for the accretion rate thus takes the form 
\begin{align}
\Mdot \sim \beta\,\frac{\M}{\tau},
\end{align}
where $\tau$ is the age of the system. Although highly approximate, this prescription is sufficient for our purposes and captures the necessary features of the accretion process.

\paragraph{Dynamo Generation} To quantify magnetic field generation in young stars and planets, we adopt the \cite{Christensen2009} scaling law, which connects the surface field strength to the intrinsic luminosity of the object. As a physical model, this relation is appropriate for rapidly rotating, fully convective astrophysical bodies, and effectively relates the magnetic energy density to the convective (kinetic) energy density, which in turn determines the heat flux, $q=\sigma\,T_{\rm{eff}}^4$. The result can be expressed in the form
\begin{align}
\frac{\langle B\rangle^2}{2\,\mu_0} = c\,f_{\rm{ohm}}\,\langle \rho \rangle^{1/3}\,(\mathcal{F}\,q)^{2/3},
\end{align}
where $c\approx 0.63$ is a constant of proportionality, $f_{\rm{ohm}}$ is the ratio of Ohmic to total dissipation, and $\mathcal{F}$ is a numerical factor that encapsulates the radial dependence of the flux, density, as well as the ratio of size of the largest convection cell to the temperature scale-height.

For young stars, \cite{Christensen2009} estimate that $f_{\rm{ohm}} \approx 0.5$ constitutes a reasonable lower bound, and demonstrate that $\mathcal{F}\sim1$ for a broad range of bodies, including Jupiter and fully convective M and K dwarfs\footnote{In particular, \cite{Christensen2009} obtain $\mathcal{F}=0.69-1.22$ for stars with masses in the range of $\M=0.25-0.7\,M_{\odot}$ and calculate that $\mathcal{F}=1.19$ for present-day Jupiter.}. In addition, \cite{Christensen2009} find that the surface field is smaller than the average interior field by a factor of $\gamma=\langle B\rangle/\B\approx3.5$. We adopt these fiducial values here for the sake of definiteness.

\paragraph{Gravitational Contraction} Young stars and giant planets contract by radiating away gravitational energy. This contraction occurs on the Kelvin-Helmholtz timescale, which can be determined by equating the bolometric luminosity, $4\,\pi\,\R^2\,q$, to the energy loss rate, $b\,\dot{\R}\,\G\,\M^2/\R^2$. Integrating the resulting relation in time, we obtain the well-known result 
\begin{align}
\R \approx \bigg( \frac{b\,G\,\M^2}{12\,\pi \, q\, \tau} \bigg)^{1/3},
\label{KHcontraction}
\end{align}
where we have dropped the surface term (which depends only weakly on initial conditions). The dimensionless constant, $b$, is determined by the interior structure of the central body. Here we take $b=3/7$, which corresponds to a $n=3/2$ (fully convective) polytrope \citep{Chandrasekhar1939}. 

A final specification that remains to be made is the mass-radius relationship of the central body. Although no exact analytic results apply over the full range of masses considered here, detailed pre-main sequence calculations of \cite{Baraffe2015} indicate that the radius scales with mass according to the approximate relation $\R \propto \M^{1/3}$, which corresponds to constant mean density. This effective mass-radius relation provides an adequate match to the evolutionary models at early times (as an example, for time $t=2\,$Myr the corresponding isochrones are best fit by a power-law of the form $\R \propto \M^{0.37}$). In fact, this result is consistent with expectations: the often-quoted ``canonical" radius of a solar-mass T-Tauri star is $\R \sim 2\,R_{\odot}$, while core-accretion calculations give $\R \sim 2\,R_{\jupiter}$ as an estimate for Jupiter's radius during the nebular epoch (e.g., \citealt{Lissauer2009,Berardo2017}). Both of these values translate to a mean density of $\langle \rho \rangle \approx 0.15\,$g/cm$^3$, which we adopt as a fiducial estimate. Combining this mass-radius relation with equation (\ref{KHcontraction}), we obtain a linear scaling of the flux on the mass and inverse dependence on the system age, i.e., $q\sim (b/9)\,\G\,\M\,\langle \rho \rangle/\tau$.

\paragraph{Result}
Substituting the relationships derived above into equation (\ref{omegatr}), we obtain an expression for the magnetospheric truncation frequency of astrophysical disks, 
\begin{align}
\Omega &= 2\, \xi \, \Bigg[ \frac{\sqrt{2}}{\big(3\,b\,\mathcal{F}\big)^2} \bigg( \frac{\pi\,\beta\,\gamma^2}{c\,f_{\rm{ohm}}} \bigg)^3 \frac{\big(\G\,\langle \rho \rangle \big)^3}{\tau} \Bigg]^{1/7} \nonumber \\
&\approx 2.4 \times 10^{-5} \, \mathrm{s}^{-1} \approx \frac{2\,\pi}{3\,\rm{day}}.
\label{result}
\end{align}
This expression provides the key result of the paper. Our analysis shows that basic considerations of gravitational stability of astrophysical disks, convective dynamo generation, and Kelvin-Helmholtz contraction translate to a disk truncation orbital frequency that is largely independent of mass. Moreover, evaluation of this result at time $\tau\sim3\,$Myr --  corresponding to the typical disk lifetime -- yields a value of the orbital period $2\pi/\Omega\approx3$ days. 

Taking into account the standard results of planetary migration theory, which predict that the equilibrium planetary orbital radius will exceed the disk truncation radius by a modest factor of $\sim1.3$ (e.g., \citealt{ataiee2021}), and considering our adoption of the lower-bound value of $f_{\rm{ohm}} = 0.5$ in our fiducial analysis, we expect the planetary period to be slightly greater. As a result, it is reasonable to anticipate that the planetary period will surpass the disk truncation period by a factor of approximately $1.5 - 2$. Furthermore, intrinsic variation in system parameters introduce a spread in the orbital periods, resulting in a range spanning a factor of approximately 2 to 3, as discussed in the Appendix. This factor of $\sim3$ variance is roughly consistent with the observed distributions of inner periods, as shown in Figure \ref{fig:data} and in Figure \ref{fig:panels} (which shows the distributions separately for each spectral type of the host). Equation (\ref{result}) thus provides a theoretical explanation for why the inner-most objects within typical extrasolar planetary systems as well as systems of giant planet satellites orbit at a period of approximately $P\sim 2-12$ days, and why this quantity is essentially independent of the mass of the central body.


Though one could consider distinct parameterizations from those adopted here, it is important to note that the characteristic orbital period varies extremely slowly as a function of both time and the physical properties of the central object. That is, at any given time, the benchmark period scales according to  $\Omega\propto\langle\rho\rangle^{3/7}$, so that moderate departures from the assumed mass-radius relation ($\R \propto \M^{1/3}$) result in only weak dependence on the host mass. In addition, as systems age, the mean density is expected to increase, but this dependence is partially compensated by the increase in time scale $\tau$ (and the entire factor in square brackets is taken to the 1/7 power). As a result, while intrinsic variation of parameters is expected to introduce some degree of variability in the period from system to system, under the assumptions that the relevant variables are independent from each other, and the fluctuations in the individual variables themselves are weak functions of the mass of the central body, the predicted boundary range of the orbital period should remain robustly mass-independent.

\section{Discussion}

The preservation of dimensionless numbers across a broad range of scales is a common phenomenon in physical systems \citep{Buckingham1914}. The near-preservation of a dimensional number, such as a frequency, is a much rarer occurrence. The theoretical analysis outlined in this work demonstrates that the characteristic orbital period associated with magnetospheric disk truncation - and by extension, planetary orbits - systematically evaluates to a few days, over a wide range of the host object's mass. Our result explains why the innermost orbital periods of planets around FGKM stars, and potentially that of large natural satellites of Jupiter and Saturn, generally fall within $\sim2-12$ days, suggesting nearly universal behavior over a wide range of host masses. One intriguing implication of this result is that despite their striking diversity, the terminal assembly of orbital architectures of planetary and satellites systems is governed by the same fundamental physical mechanism.

While our analysis provides a theoretical explanation for the characteristic orbital period of the inner-most objects in planetary and satellite systems, it is crucial to recognize the limitations of our current sample size. Presently, our knowledge of planetary systems primarily revolves around host stars with masses ranging from $\M \sim 0.1-1M_{\odot}$ \citep{Thompson2018}, and our assessment of relevant satellite systems is constrained to Jupiter and Saturn. As the census of exo-moons is expected to expand in the coming years \citep{Limbach2023, Ruffio2023}, our calculations robustly predict a characteristic orbital period for the innermost objects within these systems. The largely uncharted brown-dwarf territory, which spans masses in the range of $0.01-0.1 M_{\odot}$, presents an important additional opportunity for further exploration: our model predicts that if planetary systems that encircle brown dwarves exist, they should exhibit a similar characteristic orbital period of a few days, mirroring the trends observed in their lower- and higher-mass counterparts. Future observations of such systems will provide a valuable test of our predictions and improve our understanding of the processes that govern the formation and evolution of planetary systems.

As important as the data itself, it is crucial to consider the limitation of our model's range of applicability, which manifests primarily as bounds on the central body's mass. At the lower-mass end of the spectrum, our model is unlikely to be applicable to sub-Jovian objects, as large-scale numerical simulations \citep[e.g.][]{Lambrechts2019} suggest that Neptune-class planets do not form gaseous circumplanetary disks during the nebular epoch. Conversely, at the higher-mass end, our model is expected to falter for stars with masses exceeding $\M \gtrsim 1.5-2 M_{\odot}$. This limitation arises because these stars develop radiative cores while the gaseous disk persists \citep{Baraffe2015}, undermining the assumption of a fully convective interior that leads to a dipolar magnetic field throughout the T-Tauri phase\footnote{We remark that in principle, Jovian-class objects can also develop radiative regions within their interiors at early epochs \citep{Berardo2017}, potentially introducing similar complexity into the planetary sample.}. Toward this end, \citet{Gregory2012} have already shown that the magnetic fields of $\M < 1.5\ M_{\odot}$ stars remain dipolar for millions of years, while the fields of $\M > 2\ M_{\odot}$ stars transition to a more complex multipolar structure over the same timescale, leading to somewhat smaller truncation periods. Although the current census of sub-Jovian planets around massive stars is too limited to provide a comprehensive test of this prediction, we may reasonably expect the derived period uniformity to vanish for sufficiently massive stars.


We conclude by acknowledging that, as with any analytic theory, our results are imperfect. As already mentioned above, intrinsic variations in disk accretion rates, dynamo efficiency, and uncertainties in our description of the scaling between mass and radius, as well as bolometric flux, inevitably introduce some degree of scatter into our result. It is further worth noting that the scenario proposed in our work is only relevant for planets and satellites with enough mass to undergo significant orbital migration during the lifetime of their birth nebulae. Additionally, post-nebular processes such as tidal migration and instabilities in multi-planet systems (see \citealt{GB2021,GB2022} and the references therein) may slightly modify our predictions. Nonetheless, our framework offers a qualitative foundation for unveiling the universal role played by magnetospheric cavities in shaping the outlines of the galactic planetary census. Future observations and refinements of our theoretical framework will deepen this understanding.

\acknowledgements

We are thankful to Erik Petigura, Ravit Helled, and Andrew Howard for insightful discussions. We thank the anonymous referees for providing careful and insightful reviews of the manuscript. K.B. is grateful to Caltech, the David and Lucile Packard Foundation, and the National Science Foundation (grant number: AST 2109276) for their generous support. F.C.A. is supported in part by the University of Michigan and the Leinweber Center for Theoretical Physics. J.C.B. is grateful to Caltech and the Heising-Simons foundation for their support through the 51 Pegasi b fellowship.

\appendix
\label{sec:distrib}

\renewcommand{\thefigure}{A\arabic{figure}}
\setcounter{figure}{0}

This Appendix considers the distribution of orbital periods for the innermost planets, as set by magnetic truncation. Equation (6) in the text gives the orbital frequency corresponding to the truncation radius, and shows that the period is about 3 days for typical values of the input variables.  To estimate the distribution of values, we note that the expression for the period can be written as the product of variables. These variables (the field geometry factor $\xi$, the binding energy factor $b$, and so on) can vary from system to system, sampling some probability distribution, thereby resulting in a range of inner periods. In the limit where the number of variables is large, the distribution of the composite variable (the period) would approach a log-normal form. Even with a finite number of variables, however, we can estimate the width of the composite distribution with some degree of certainty (see \citealt{fatadams1996} for an analogous calculation).

We start by writing the expression for the period in the form 
\be
P = P_0 \xi^{-1} \xi_m^{-3/7} \xi_\rho^{-3/7} \xi_b^{2/7} \,,
\label{prodone}
\ee
where $P_0$ is the fiducial value of about 3 days, and the variable $\xi$ is the same as before. We have also defined 
\be
\xi_\rho = {\rhobar \over \rhobar_0} \qquad {\rm and}
\qquad \xi_b = {b \over b_0}
\ee
and we have collected all of the magnetic parameters into a single dimensionless quantity
\be
\xi_m = {\beta\over\beta_0} {\gamma^2\over\gamma_0^2} 
\left({c \over c_0} {f_{\rm{ohm}}\over f_{\rm{ohm\,}0}}\right)^{-1} 
\left( {{\cal F} \over {\cal F}_0} \right)^{-2/3} \,.
\ee
Note that we could in principle include each individual variable. However, as discussed below, the distributions are not known, so that is it simpler to use the composite `magnetic variable' $\xi_m$ defined here. Moreover, the input variables that make up this composite variable might not be fully independent. Taking the logarithm of equation (\ref{prodone}) we find
\be
\log(P/P_0) = \sum_k p_k \log\xi_k\,,
\ee
where the index runs from 1 through 4 and the exponents $p_k$
are those given in original equation (\ref{prodone}).

\begin{figure}
    \centering
    \includegraphics[width=0.5\linewidth]{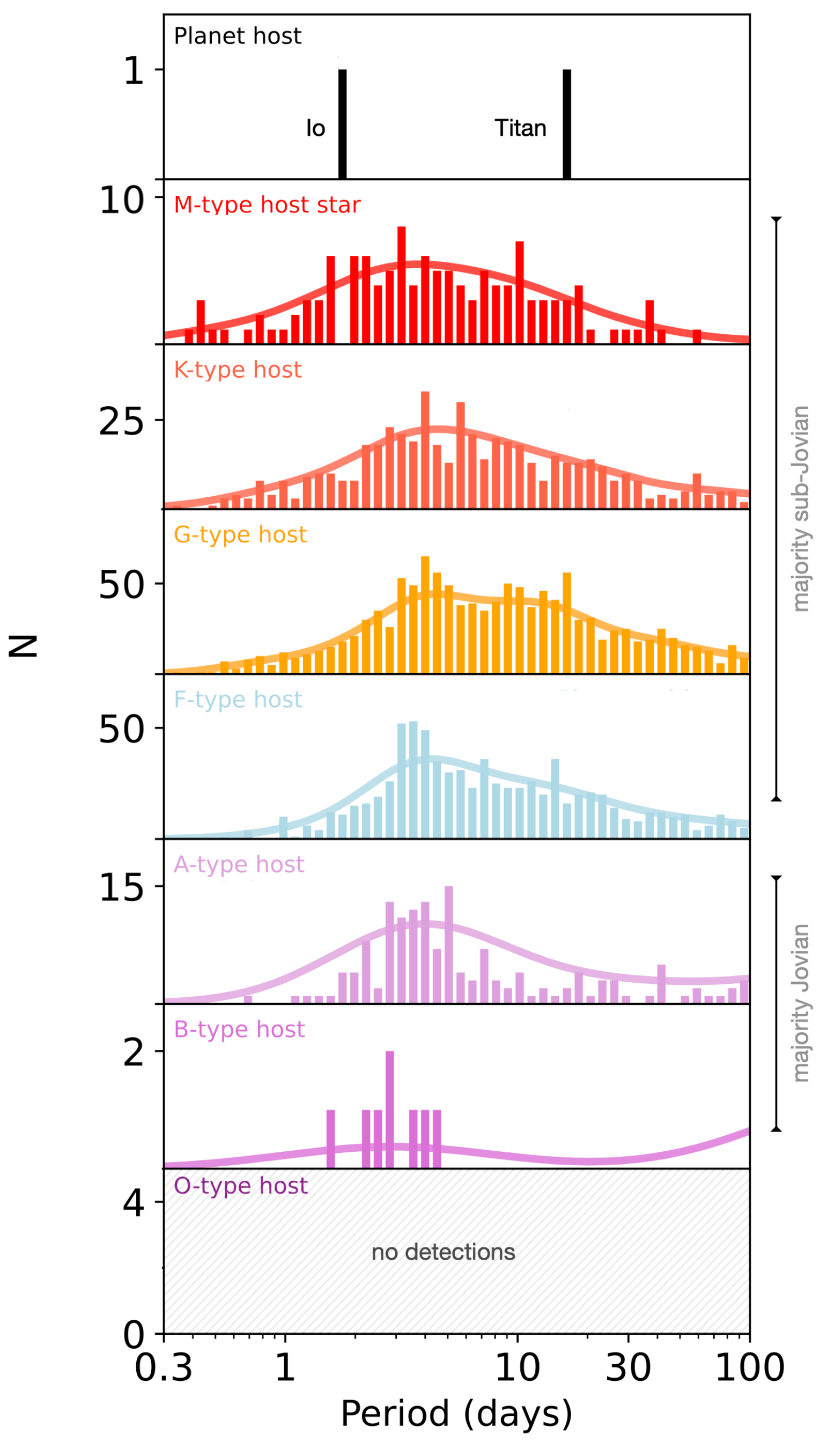}
    \caption{Distribution of innermost orbital periods, sorted by host type. Unlike in Figure \ref{fig:data}, here we do not restrict the data-set to systems with measured masses, yielding a somewhat distinct view of the galactic planetary census. The orbital periods of Jupiter's moon Io and Saturn's moon Titan are also shown for comparison. It is important to note that both bodies have experienced some degree of outward tidal migration. In the case of Titan, the work of \citet{Lainey2020} suggests that long-range tidal migration might have occurred. The distributions for M, K, G, and F-type hosts appear quite similar. For A and B-type hosts, there is a slight preference for lower periods. However, it's worth noting the planetary sample for these types of stars is different: while the fractions of planets with radii smaller than 4 Earth radii are 56\%, 66\%, 69\% and 45\% for M, K, G, and F-type stars, respectively, it is 14\% and 0\% for A and B-type stars, illustrating the sample disparity.}
    \label{fig:panels}
\end{figure}

Each random variable $\xi$ has the form $\xi = x/x_0$. As a result, the mean of the logarithm of each variable is given by 
\be
\langle \log\xi \rangle = \langle \log x \rangle -
\log x_0 \,,
\ee
where angular brackets denote taking the average over the distribution of the variable.  If we take the fiducial values $x_0$ of the physical variables to be defined such that $\log x_0$ = $\langle\log{x}\rangle$, then the logarithm of the random variables $\xi$ will have zero mean $\langle\log\xi\rangle$ = 0. Keep in mind that $x_0$ is not the mean value itself, but rather $x_0$ = $\exp[\langle\log{x}\rangle]$ $\ne\langle{x}\rangle$.

If we then assume that the variables are independent, the variance of the quantity $\log(P/P_0)$, which is a sum of random variables, is given by the sum  
\be 
\sigma_T^2 = \sum_k p_k^2 \sigma_k^2\,,
\ee
where $\sigma_k^2$ is the variance of the variable
$\log\xi_k$.

To fix ideas, consider the case where each physical variable $x_k$ can vary by a factor of $\Lambda_k$ from its fiducial value. In the simplest case, the corresponding variable $\log\xi_k$ will thus have a uniform distribution over the range $[-\log\Lambda_k,\log\Lambda_k]$. The variance of the individual distribution is given by  
\be
\sigma_k^2 = {2\over3} (\log\Lambda_k)^2\,.
\ee
The total variance of the period distribution, the composite
variable $\log(P/P_0)$ then has the form 
\be
\sigma_T^2 = {2\over3} \sum_k p_k^2 (\log\Lambda_k)^2\,.
\ee
Although we do not know the distributions of the variables in question, they are expected to vary by modest factors. For the sake of definiteness, let's assume that $\Lambda_k$ have the same value (denoted now as just $\Lambda$) for all variables/indices $k$. Our expression becomes 
\be
\sigma_T^2 = {2\over3}(\log\Lambda)^2 \sum_k p_k^2
= {142\over 147} (\log\Lambda)^2 \approx(\log\Lambda)^2 \,. 
\ee
Recall also that $\sigma_T^2$ is the variance in $\log(P/P_0)$.  As a result, if each individual variable $x_k$ can vary by a factor of $\Lambda$, then the period will vary by a factor of $\sim\Lambda$ also. The increase in variance that arises from the multiple random variables is offset by the small exponents.  Since the individual variables are expected to vary by factors of 2 to 3, the predicted period should lie in a range of about 2 to 12 days. Although not negligible, this factor of $\sim3$ is small compared to the range of masses (a factor of $\sim1000$) for the central bodies.



\bibliography{references}{}
\bibliographystyle{aasjournal}

\end{document}